\documentstyle[12pt,epsf]{article}
\makeatletter
\@addtoreset{equation}{section}

\def\baselinestretch{1.2}
\def\href#1#2{#2}

\newcommand{\be}{\begin{equation}}
\newcommand{\ee}{\end{equation}}
\newcommand{\beq}{\begin{eqnarray}}
\newcommand{\eeq}{\end{eqnarray}}
\newcommand{\ba}{\begin{array}}
\newcommand{\ea}{\end{array}}

\renewcommand{\a}{\alpha}

\renewcommand{\d}{\delta}

\newcommand{\e}{\varepsilon}

\newcommand{\th}{\theta}

\newcommand{\nl}{\hspace{-.65cm}}

\begin{document}

\begin{titlepage}
\newcommand{\preprint}[1]{\begin{table}[t]  
           \begin{flushright}               
           \begin{large}{#1}\end{large}     
           \end{flushright}                 
           \end{table}}                     
\preprint{hep-th/0211140\\PUPT-2060}

\begin{center}
\LARGE{Anyons, 't Hooft loops  and a Generalized\\ Connection
in Three Dimensions}

\vspace{10mm}

\normalsize{N. Itzhaki}\vspace{10mm}

{\em  Department of Physics, Princeton University,
Princeton, NJ 08544, USA}\end{center}\vspace{10mm}

\begin{abstract}

We consider a generalized connection in three dimensions and show that
it emerges in Chern-Simons-Maxwell theories when one studies
the disorder instanton operator.
We generalize this construction to non-Abelian theories and find that the disorder operator
(the 't Hooft operator) is equivalent to a generalized Wilson loop in a representation that
depends on the Chern-Simons term. We speculate about the effective action of
the disorder operator and its applications to the possible phases of the theory in the infra-red.
We also show that fractional statistics can emerge in gauge theories
without a Chern-Simons term if the generalized connection rather than the ordinary
connection is used to couple charged particles.

\end{abstract}
\end{titlepage}
\newpage
\baselineskip 18pt
\renewcommand{\baselinestretch}{1.05}  

\section{Introduction}

One of the most interesting aspects of physics in three dimensions is that the linking number
is well defined. As a result  fractional statistics can emerge in 3D. In gauge theories this is usually realized
 by adding a Chern-Simons term to the action. In this paper we show that 3D gauge theories are special
in another, closely related, respect; they admit a generalized connection and a generalized Wilson loop.
As it turns out with the help of the generalized connection one can realize fractional statistics without
 introducing  a Chern-Simons term. This provides a novel way to describe anyons.

The generalized Wilson line appears also in a somewhat unexpected place. In
Maxwell theory the disorder operator is a singular instanton that can be thought of as
the exponent of the dual scalar.
The creation operator of such an instanton can be
written explicitly in terms of the field strength. When one generalizes this to a Chern-Simons-Maxwell
theory one finds that the disorder operator is now an open generalized Wilson line. Probably
the most interesting aspect of this result is that it can be generalized to the non-Abelian $SU(N)$ case.
In that case the disorder operator is a $Z_N$ instanton, known as the 't Hooft point \cite{thooft},
that appears at the end point of a $Z_N$ Dirac particle. This operator plays a crucial role in determining
the phase of the theory in the infra-red. We show that the 't Hooft operator is {\em equivalent} to
a generalized Wilson line in a representation that depends on the Chern-Simons term.

The paper is organized as follows. In section 2 we describe the generalized connection and discuss its physical meaning.
The precise relation with anyons and fractional statistics is made in section 3. In section 4 we show that
in Chern-Simons-Maxwell theory the disorder parameter is described by an open generalized Wilson line.
The interesting fact that this construction can be generalized to the non-Abelian theory is discussed in section 5.
Section 6 deals with the effective description of the disorder operator.

\section{A generalized connection}

In three dimensional gauge theories one can define a generalized
connection in the following way
\be
A_{\mu}^{\th}\equiv A_{\mu}+\frac{\th}{ g^2} F_{\mu}, \;\;\;\mbox{where}\;\;\;
F_{\mu}=\frac12\e_{\mu\nu\rho} F^{\nu\rho}.
\ee
$g$ is the coupling constant which in 3D has dimension $1/2$ and
$\th$ is a dimensionless parameter that is real in Minkowski space-time and is imaginary
with Euclidean signature.
Under a gauge transformation
$A_{\mu}^{\th}$ transforms like a connection for any value of $\th$\footnote{In
non-Abelian theories one can consider  a more generalized connection
by taking $\theta$ to be  an hermitian matrix. In $U(N)$ theories this has the effect of mixing between
the $SU(N)$ and the $U(1)$ which we would like to avoid in this paper. We should  however
mention that such a connection might be interesting since it gives a new realization of
 non-Abelian anyons. In the present paper we consider only Abelian anyons (even in non-Abelian gauge theories).}
\be
A^{\th}_{\mu} \rightarrow g^{-1}(x) A^{\th}_{\mu} g(x) + i g^{-1}(x) \partial_{\mu} g (x).
\ee
We can consider, therefore,
a new set of non-local gauge invariant operators
that are natural generalizations of the Wilson loops
\be
W^{\th}=\mbox{Tr}P\exp(i\oint dx^{\mu} A_{\mu}^{\th}).
\ee
One might wonder whether  $W^{\th}$ makes sense at the quantum level.
The basic concern is that since $W^{\th}$ involve both the gauge fields and
their canonical conjugate fields, their order must be specified for the operator to be well defined.
However, from the commutation relation between the gauge field and the field strength it follows
that $[A_i, F_i]=0$ ($i=1,2$)
and hence there is no quantum
ambiguity in the definition of $W^{\th}$.
This also implies that, operators in which $\th$ varies along the loop are also consistent.
On the other hand $A_{1}^{\th}$ and
$A_{2}^{\th}$  do not commute for $\th\neq 0$.
Therefore the generalized Wilson loop is a consistent operator at the quantum
level only in the absence of intersections. This is reminiscent of ordinary Wilson loop in
Chern-Simons theories where the fact that $A_{1}$ does not commute with $A_{2}$ implies
that a Wilson loop is well defined only without intersections. As we shall see there are other
similarities between the generalized Wilson lines in Maxwell theory and  ordinary Wilson lines
in CS theories.

Roughly speaking, in Abelian theories, the physical meaning of an ordinary
Wilson loop is that it creates an electric charge along the loop
and it measures the magnetic flux that is passing through the
surface bounded by the loop. What is the physical meaning of the generalized Wilson loop?
With the help of  Stokes's theorem
and the equations of motion, $\partial_{\mu}F^{\mu\nu}=j^{\nu}$, the
generalized Wilson line can be
written as a surface integral,
\be\label{2}
W^{\th}_P=\exp(i\oint_P dx^{\mu} A_{\mu}^{\th})=
\exp i(\Phi_M+\th Q) ,
\ee
where $\Phi_M$ is the total magnetic flux
passing trough the surface and $Q$ is the total electric
charge. We see that the generalized Wilson loop
measures the total electric charge times $\th$  plus the magnetic
flux. Since
$Q$ in an integer $W^{\th}$ is a function only of $\th$  mod $2\pi$.
That statement is correct only in the Abelian theory.
To find what the generalized Wilson loop is creating we
 add $\log(W^{\th})$ as a source to the action
and solve the equation of motion. In the Abelian case this is an easy task.
Consider for simplicity a
generalized Wilson line  along a straight line in the $x_0$ direction. We add to the Maxwell
action $\int_{-\infty}^{\infty}dx_0A_0^{\th}$ and solve the equations of motion
\be\label{p2}
A_0=\frac{g^2}{2\pi}\log (r),\;\;\;\;\;A_1=\frac{\th
x_2}{2\pi r^2},\;\;\;\;\;A_2=- \frac{\th x_1}{2\pi r^2}, \;\; \;\;\;
 r^2=x_1^2+x_2^2.
\ee
This background describes one unit of an electrically charge particle and an infinitely
 thin magnetic flux with total flux $\th$ both located at the origin.
So the generalized Wilson loop creates a bound state of an electric charge and a magnetic flux.
Such a bound state is known as an anyon.

Now consider the following generalized Wilson loop $W^{\tilde{\th}}$
that is winding around the origin  and
is probing the background (\ref{p2}).


\[
\parbox{6.0in}{
\begin{picture}(150,120)(0,0)
\vspace{5mm} \hspace{20mm} \mbox{\epsfxsize=100mm \epsfbox{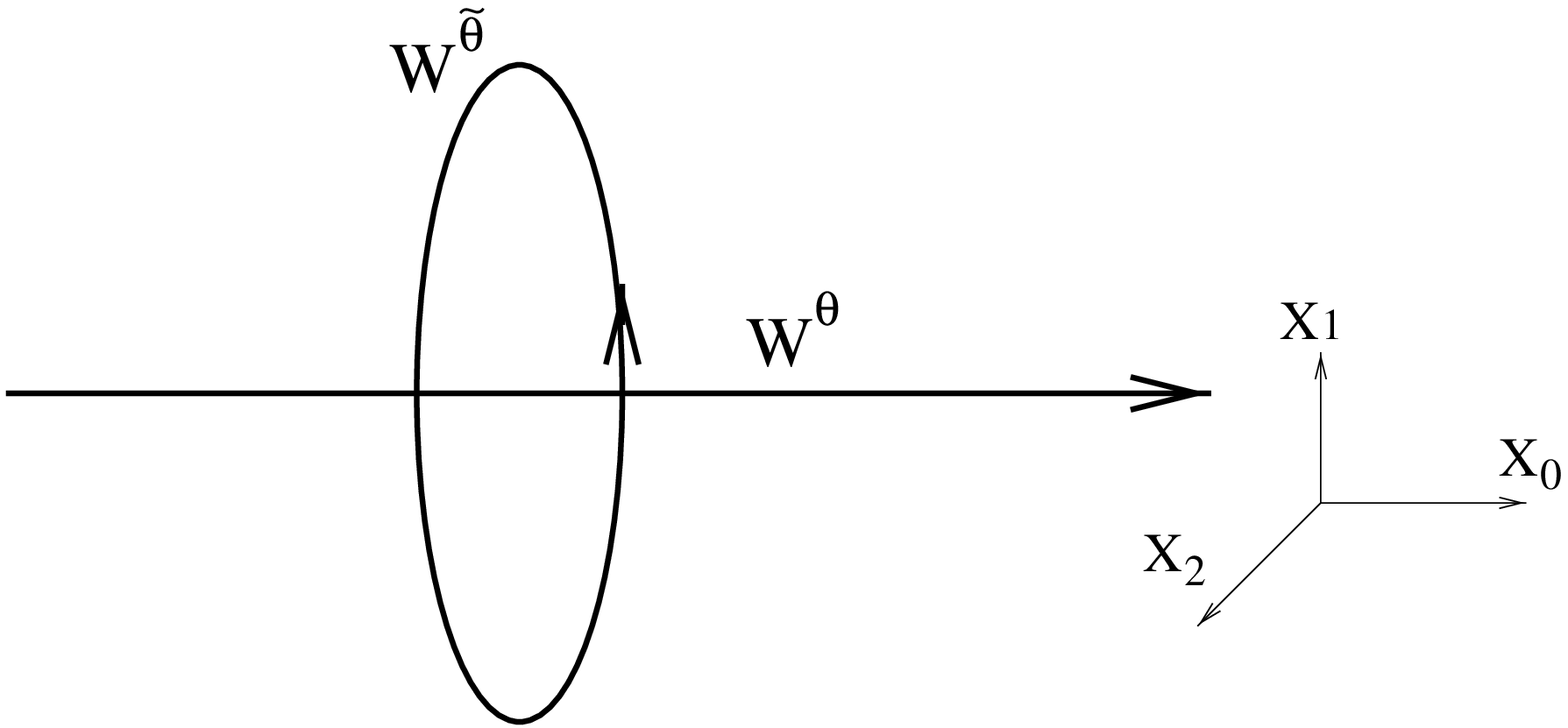}}
\end{picture}}
\]
From eqs.(\ref{2},\ref{p2}) we find that such a loop will pick up a phase
\be\label{1}
\exp(i n(\th + \tilde{\th})),
\ee
where $n$ is the winding number.
 It is easy to show that (\ref{1}) holds for general loops
$P_1$ and $P_2$ where  $n$ is the linking number. Note that the  phase is not trivial
for  $\tilde{\th}=\th$.
 Again the similarity with ordinary
Wilson lines in Chern-Simons theory \cite{Polyakov,Witten:88} shows up.
The origin of that similarity can be traced to the fact that
\be\label{2op}
\frac{k}{2\pi} \langle A_{\mu}(x) A_{\nu} (y) \rangle_{CS}=
\frac{1}{g^2}\langle A_{\mu}(x), F_{\nu}(y)\rangle_{Maxwell} = i\frac{\e_{\mu\nu\rho}}{4\pi}
 \frac{(x-y)_{\rho}}{(x-y)^3},
\ee
and hence upon integrating the exponent of (\ref{2op}) over $P_1$ and $P_2$  one gets the
Gauss linking number
\be
\frac{i}{4\pi}\oint_{P_1}\oint_{P_2}dx^{\mu} dy^{\nu} \e_{\mu\nu\rho}
 \frac{(x-y)_{\rho}}{(x-y)^3}=n(P_1,P_2),
\ee
that is known to be
an integer as long as $P_1$ and $P_2$ do not intersect.
In the next section we elaborate further on this relation.

\section{Anyons}

As it is well known  physics in three dimensions is special since it allows
for the possibility of particles with any statistics \cite{wilczek}.
The reason is that in three dimensions the number of times that a
particle winds around  another particle as the two evolve from past infinity
to future infinity is  well defined.
Therefore, it is meaningful to view particles interchange as
"half-winding" and to determine
their statistics to be one-half of the phase that is picked up when one particle winds once
around the other.\footnote{\label{fo}To be more precise the statistics is
determined in that way only up to a minus sign. For example, there are two ways to realize
 no-phase: $1^2=1$ that corresponds to bosons, and $(-1)^2=1$ that corresponds to fermions.
In other words, the fact that in pure Maxwell theory in three dimensions an ordinary Wilson loop does not
pick up a topological phase while winding around another ordinary Wilson loop does not tell us whether the source for
the Wilson loop is a charged fermion or a charged boson. }
A concrete way to realize non-trivial statistics is by attaching a magnetic flux, $\Phi$, to
electrically charge particles. The Aharonov-Bohm effect then implies a  shift,
 that is linear in $\Phi$, in the
statistics of the particles. The question is how to
 attach a magnetic flux to the electrons.
The simplest way considered so far
is via a Chern-Simons theory. As the discussion in the
previous section suggests exactly this effect can be achieved also in the
context of pure Maxwell theory without a Chern-Simons term but
with a generalized connection.\footnote{A different way to realize fractional statistics that does not involve
CS theory was discussed in \cite{11}.}
 Before we elaborate on this let us
review the CS construction.
The relevant Lagrangian is
\be\label{q4}
{\cal L}(\a)=j^{\mu} A_{\mu} -\frac{\a}{2}\epsilon^{\nu\rho\sigma}
A_{\nu}\partial_{\rho} A_{\sigma},
\ee
where $j^{\mu}$ is the current. The first term is
the standard current gauge field coupling the second term is the Abelian Chern-Simons term.
We do not write down the pure matter field term as it plays no
role here.
From the zero component of the equation of motion for $A_{\mu}$ one finds
that  the total number of charged particles $N$ and the total magnetic flux are related
\be\label{q1}
N=\a \Phi.
\ee
This  implies that  a magnetic flux is attached to every charged particle.
The shift in the  naive statistics of the particle is fixed by the Aharonov-Bohm effect to be
\be\label{k1}
\Delta\phi=1/2\a.
\ee
Notice that because of the CS term in (\ref{q4})  the phase
(\ref{k1}) is only one-half the  Aharonov-Bohm phase.

We would like to show now that eq.(\ref{q1}) can be obtained without a
Chern-Simons term but with the help of the generalized connection.
 Consider the following Lagrangian
\be\label{q2}
{\cal L}(\th)
=-\frac{1}{4g^2} F^2+  j^{\mu} A^{\th}_{\mu},
\ee
which describes   Maxwell theory that couples to
the current via the generalized connection rather then the usual connection.
Note that this coupling is gauge invariant as long as $j^{\mu}$ is a conserved current.
Notice further  that the
 $ \epsilon_{\mu\rho\sigma}$ in the generalized connection leads to
 time reversal and parity violation which is crucial to realize fractional statistics.
The equation of motions for $A_{\mu}$ give
\be\label{q3}
\Box A_{\mu}=g^2 J_{\mu}+ \th\epsilon_{\mu\nu\rho} \partial_{\nu}J^{\rho}.
\ee
Consider now   a static point like particle located at the
origin. Such a particle is described by $j^{0}=\d (x)$. Plugging this into
eq.(\ref{q3}) and solving for the gauge fields one finds (\ref{p2}) which implies
that  the magnetic flux  attached to $N$ electrons is
\be\label{k2}
N=\frac{\Phi}{\th},
\ee
and that the shift in the statistics of the particle is
\be\label{k3}
\Delta \phi =\th.
\ee
Eqs.(\ref{k2},\ref{k3}) differ from eqs. (\ref{q1},\ref{k1}) by a
 factor of $2$. The reason is that with the generalized connection construction a long range
electric field is also generated (see eq.(\ref{p2})) which couples to the current and gives
exactly the same phase. Simply put in the CS case the phase is due to a charged particle winding
around a magnetic flux while here we also have the contribution of a flux tube winding around the
charged particle.

It is appropriate, therefore, to call   the generalized connection an anyonic
connection for it generates the parallel transport of an anyon. This can
be phrased in terms of gauge invariant operators.
The gauge invariant operator that creates an anyon at some point
$x$ propagates it along a certain path to point $y$ where it is
annihilated is
\be
\Psi(x) P \left\{ \exp(i \int_{x}^y dx^{\mu}A^{\th}_{\mu})\right\}\bar{\Psi}(y).
\ee

Although both ${\cal L}(\a)$ and
${\cal L}(\th)$ lead to fractional statistics by the same mechanism of attaching a magnetic flux
 to the electrons
the physics they describe is quite different.
${\cal L}(\a)$ is a super-renormalizable theory
while, at least naive power counting suggests that, ${\cal L}(\th)$ is a non-renormalizable
 theory that makes sense at the quantum level
only with a cutoff. The infra-red is also  different.
 In ${\cal L}(\a)$ there are no dynamical
 degrees of freedom associated with
the gauge fields. The sole function of the gauge fields is to attach a magnetic flux
to the electrons.
On the other hand there certainly are  gauge degrees of freedom in
${\cal L}(\th)$.  Moreover, in the absence of a CS term the gauge fields are massless
 and can widely fluctuate in the infra-red.
If one's goal is to suppress these fluctuations
 then one can, for example, consider a CS theory  that couples
to the current via the generalized connection rather then the usual connection.
 This has the interesting effect that the previously singular
magnetic flux tube is smeared.
As a result the mean field theory approximation used to study anyon superconductivity \cite{lau}
is very likely to be under  better control, possibly even away from the almost fermionic limit
studied in detail in \cite{ww}.

\section{Disorder operator in Maxwell-Chern-Simons theory}

In Maxwell theory (without a Chern-Simons term)  one can
define a {\em local} disorder operator
\be\label{lkj}
\eta(x)=\exp\left( \frac{2\pi i}{g^2}\int_{x}^{\infty}dx^{\mu} F_{\mu}\right) .
\ee
To show that this is  a local operator one needs to show that $\eta(x)$ does not
depend on the path from $x$ to $\infty$. In other words a closed loop should be trivial.
 With the help of Stokes's
theorem and the equations of motion
 we find that
\be
\exp\left(\frac{2\pi i}{g^2}\oint dx^{\mu} F_{\mu}\right)=\exp(2\pi i Q)=1,
\ee
where $Q$ is the total charge passing through the surface.
 $\eta(x)$  is the creation operator of a Dirac magnetic instanton
 at $x$ with a Dirac particle (the three dimensional analog of the Dirac string)  along
the path from $x$ to infinity. The statement that this is a local operator
 is nothing but the statement that with the appropriate quantization condition the Dirac
particle is not a physical observable. In terms of the disorder operator the dual scalar
 is defined as
\be\label{kk1}
\phi(x)=\frac{g^2}{2\pi i} \log (\eta(x)),
\ee
which with the help of eq.(\ref{lkj}) can be written in a
 more familiar form  $\partial_{\mu}\phi= F_{\mu}$.

Let us generalize the construction of the disorder operator to  Chern-Simons-Mawxell
theory. The Lagrangian is
\be\label{rr}
{\cal L}=-\frac{1}{4 g^2} F^2 +\frac{k}{4\pi} F^{\mu}A_{\mu}.
\ee
If we use the same definition for $\eta(x)$ we find that since the equations of motion
are modified by the CS term, $\partial_{\mu} F^{\mu\nu}=j^{\nu}-\frac{k}{2\pi} F^{\nu}$,
 a closed loop gives
\be
\exp\left(\frac{2\pi i}{g^2}\oint dx^{\mu} F_{\mu}\right)=\exp\left( 2\pi i(
Q-\frac{k}{2\pi} \Phi_{B})\right) = \exp (-i k \Phi_{B}).
\ee
As $\Phi_{B}$ is not quantized this is not a trivial operator and hence $\eta(x)$ is not a local
operator.
The following simple modification  fixes that problem
\be\label{kk2}
\eta(x)=\exp\left( i\int_{x}^{\infty}dx^{\mu}( kA_{\mu}+ \frac{2\pi }{g^2}F_{\mu})\right) .
\ee
One can easily verify that with the additional term the closed loop is trivial.
Note that eq.(\ref{kk2}) can be written as a generalized Wilson line
\be
\eta(x)=\exp(i k\int_{x}^{\infty}dx^{\mu} A_{\mu}^{\frac{2\pi}{k}}),
\ee
which suggests that, as we shall see in the next section,
 a non-Abelian generalization should be fairly straight forward.

An interesting  by product of this discussion is the following.
Typically in gauge theories a charged field is not gauge invariant. Therefore, if we want
to describe the creation operator of an electron at some point $x$ in a gauge invariant fashion
we must attach a  Wilson line  to the field $
\Psi(x)\exp(i \int_{x}^{\infty} dx^{\mu}A_{\mu}).$
That operator, however, is non-local as it depends on the path. So one can either work
with a non gauge invariant operator or with a gauge invariant operator that is not unique.
In a Chern-Simons theory with $k=1$ one can define a  gauge invariant operator
that is also local $\Psi(x)  \exp(i \int_{x}^{\infty} dx^{\mu}A^{2\pi}_{\mu}).$
Note that this operator does not create  an electron but rather an electron-disorder
 bound state.

We close this section with a comment about the dual scalar in this case.
Since (\ref{kk2}) is a local operator it can be used to define
 the dual scalar. Following
(\ref{kk1}) we get
\be
\partial_{\mu} \phi= F_{\mu} + \frac{g^2k}{2\pi} A_{\mu}.
\ee
The fact that for $k\neq 0$ the connection appears in this relation
implies that $\phi$ is not gauge invariant.
In fact under a gauge transformation $A \rightarrow A +d \Lambda$ the transformation of the
dual scalar is $\phi(x) \rightarrow \phi(x) +\frac{g^2k}{2\pi} \Lambda(x)$.
Therefore, without gauge fixing the only gauge invariant action that includes only $\phi$
is a constant. With partial gauge fixing it is possible that a dual scalar action can be
constructed. For example in the Lorentz gauge $\partial_{\mu}A^{\mu}=0$ the residual gauge
symmetry is $\Box \Lambda =0$ and so $\tilde{\phi}=\Box \phi $ is a  scalar that is invariant
under the residual gauge symmetry and it can be used to construct a non-trivial  action.

\section{Disorder operator in Non-Abelian theories}

In this section we  generalize the construction of the disorder operator
 to non-Abelian Yang-Mills-Chern-Simons theories.
 Since in the  construction in the previous section we used
 Stokes's theorem  it cannot be generalized to non-Abelian theories.
A different approach  was used by Moore and Seiberg \cite{ms}
to show that in pure CS theories a 't Hooft loop is equivalent to a Wilson loop.
Below we generalize their results and show that with a Yang-Mills term a 't Hooft loop
is equivalent to a generalized Wilson loop.
First let us apply that approach to the Abelian case and rederive (\ref{kk2}).
The Maxwell-Chern-Simons action is invariant under continuous gauge transformations.
Singular gauge transformation, however, can define an operator.
Consider a closed loop, $C$,  that is parameterized by $\tau$ and perform a gauge
transformation that near the loop takes the form $\Lambda(\phi)=c\phi$ where $\phi$ is the angular
variable that winds the loop. The effect of such a singular gauge transformation
is to create a magnetic flux tube along the loop with total magnetic flux $2\pi c$.
As a result the action is not invariant under such a singular gauge transformation
\be
S \rightarrow S+ 2\pi c \oint_C dx^{\mu}( kA_{\mu}+ \frac{2\pi }{g^2}F_{\mu}),
\ee
which implies that in the path integral this singular gauge transformation is equivalent
to adding the operator
\be\label{pp1}
\exp\left( i2\pi c\oint_C dx^{\mu}( kA_{\mu}+ \frac{2\pi }{g^2}F_{\mu}) \right).
\ee
For $c=1$ (or any other integer) the singular gauge transformation cannot be detected by any
of the charged fields and hence $\eta(x)$ as defined in eq.(\ref{kk2}) is a local operator.

Consider $SU(N)$ gauge theory (or $SU(N)/Z_N$ to be
more precise) on $R^3$ and suppose that all the fields in the theory are in
 representations of the gauge group that is invariant under the center
 (for example all fields are in the adjoint representation).
In that case all fields are invariant under a gauge transformation that near the loop takes the
form
\be\label{pp2}
U(\phi)=\exp (i\frac{\phi}{N} T),~~~~~~~~T=\mbox{diag} (1,1,...,1, 1-N),
\ee
since $U(2\pi)=e^{2\pi i/N}$ is in the center. We can, therefore, follow the same steps that led to
(\ref{pp1}) and find a trivial loop
operator in $SU(N)/Z_N$ theories that in turn can be used to
construct the disorder operator (which in that case is known as the 't Hooft point \cite{thooft}).
The first step is to fix the normalization  to be the standard ones,
$\mbox{Tr}T_a T_b =\frac12 \delta_{ab}$. With that normalization the singular gauge
 transformation, eq. (\ref{pp2}), takes the form $U(\phi)=\exp(i\phi \sqrt{2(N-1)/N} H_{N-1})$
where $H_{N-1}=\sqrt{1/2(N-1)N}\mbox{diag}(1,1,...,1-N)$ is the $N-1$ generator
in the Cartan sub-algebra (we are using the notation of \cite{georgi}).
The effect of such a singular gauge transformation is to create a magnetic flux tube along the
loop that points in the $H_{N-1}$ direction. This in turn yields the following
 shift in the action
\be\label{jjh}
\delta S= k\sqrt{\frac{N-1}{2N}}  \oint_C dx^{\mu}\mbox{Tr}\left(
H_{N-1}(A_{\mu}+\frac{2\pi}{k} F_{\mu})\right) =k\sqrt{\frac{N-1}{2N}}  \oint_C dx^{\mu}\mbox{Tr}
(H_{N-1}A^{2\pi/k}_{\mu}).
\ee
Just like in the Abelian case we conclude that the singular gauge transformation along the
loop is the exponent of $\delta S$. There is, however, an important difference. Eq. (\ref{pp1})
is gauge invariant while the non-Abelian generalization, $\exp(i \delta S)$, is not
since it points in a certain direction in the gauge group.
To find the gauge invariant operator that is equivalent to the 't Hooft loop we need
 to perform an integration over smooth gauge transformation along the loop.
Here is where the fact that we can express eq.(\ref{jjh}) in terms of the
 generalized connection
is useful. The transformation of $A^{2\pi/k}_{\mu}$ under a gauge transformation
 $g(\tau)$ (recall that $\tau$ is the parameter
along the loop) is  $A^{2\pi/k}_{\mu}\rightarrow g(\tau)^{-1}A^{2\pi/k}_{\mu}g(\tau)+
ig(\tau)^{-1}\partial_{\mu}g(\tau)$ and thus averaging $\exp(i \delta S)$
over the gauge group gives
\be\label{uui}
\int Dg(\tau) \exp\left(k\sqrt{\frac{N-1}{2N}} i\oint_C dx^{\mu}\mbox{Tr}
[H_{N-1}(g(\tau)^{-1}A^{2\pi/k}_{\mu}g(\tau)+ig(\tau)^{-1}\partial_{\mu}g(\tau))] \right),
\ee
which describes  the quantum mechanics of the coadjoint orbit.
 Exactly that problem was studied in \cite{afs,dp} where it was found that
\be\label{3d}
\int Dg(\tau)\exp\left(i\oint_C dx^{\mu} \mbox{Tr}[l_i H_i(g^{-1}A_{\mu}g+ig^{-1}
\partial_{\mu} g)]\right)=\frac{1}{d(r)}\mbox{Tr}_r P \exp(i\oint_C dx^{\mu} A_{\mu}).
\ee
$H_i$ are the Cartan subalgebra generators and the vector $l_i$ is the
highest weight of the representation $r$ whose dimension is $d(r)$. Both sides of (\ref{3d})
make sense only if $l_i$ is the highest weigh of some representation.
The simplest case is $SU(2)$ where there is only one generator in the Cartan subalgebra that
can be chosen to be $\tau_3$. Then eq.(\ref{3d}) takes the form $W_{J}=
\int Dg(\tau) \exp(i J \oint dx^{\mu} \mbox{Tr}[\tau_3(g^{-1}A_{\mu}g+ig^{-1}
\partial_{\mu} g)])$ with $J=1/2, 1, 3/2,...$.

Before we apply (\ref{3d}) to the 't Hooft loop calculation
we would like to provide an argument why eq.(\ref{3d}), that at first
looks very surprising,  is actually quit intuitive. For a detailed proof see \cite{afs,dp}.
Consider a Wilson loop in $SU(2)$ Yang-Mills theory in the fundamental in arbitrary dimensions.
One way to think about
the Wilson loop is as the exponent of the phase that a quark picks up along the loop which is the right
hand side of (\ref{3d}).
An alternative way to think about the Wilson loop is the following. Let us cut a small sphere,
$S^{d-2}$, around the loop so that now there is  a boundary with topology $S^1 \times S^{d-2}$.
Integration by part implies that the boundary term is
\be
\int_{S^1\times S^{d-2}} \mbox{Tr} (A\wedge *F).
\ee
The fact that we have a charged field in the fundamental representation
 along the loop means  that $\int_{S^{d-2}} *F$ is "one" pointing in some direction
in the gauge group, say, $\tau_3$. And so upon averaging over the
gauge transformation on the loop the boundary term gives the left hand side of (\ref{3d}).

To complete the 't Hooft loop operator calculations we notice that for $k=1$
the $l_i$ in eq.(\ref{uui}) agree with the highest weight in the fundamental representation
and so we conclude that in that case a 't Hooft loop is equivalent to a generalized Wilson
loop in the fundamental representation
\be
T(C)=W_N^{2\pi}(C).
\ee
For $k>1$ we get
\be\label{456}
T(C)=W_{\mbox{sym}k}^{\frac{2\pi}{k}}(C),
\ee
where by $\mbox{sym}k$ we mean that the trace is taken in
the symmetric piece of $\overbrace{N\otimes N\otimes...\otimes N}^{k~ \mbox{times}}$ .

To define the disorder operator we need, like in the Abelian case,
to consider open 't Hooft lines. We denote by $t(x)$ the field that describes the
disorder 't Hooft point that is located at the end
point of the open 't Hooft line.
Eq. (\ref{456}) implies that $t(x)$ must be in the $\mbox{sym}k$ representation.
One interesting aspect of this result is the following. We started by defining a 't Hooft
loop which is a singular gauge transformation that cannot be detected by any of the fields
in the action.
In terms of Wilson loops what this means is that any Wilson loop, ordinary
or generalized, in a representation of the gauge group in which
the center of the gauge group acts trivially
cannot detect the 't Hooft loop. That is to say that its expectation
value is the same whether or not it is winding a  't Hooft loop. The expectation value
of  Wilson loops in other
representations that are sensitive to the center of the gauge group,
for example the fundamental representation, depends on whether or not
a 't Hooft loop is present. But we do not worry about them since there
 are no dynamical fields in these representations.
We are, therefore, tempted to claim that the 't Hooft loop is a trivial operator in such
 theories and that the 't Hooft point is a local operator.
However, we have just showed that the 't Hooft loop is equivalent to
a generalized Wilson loop that is not
necessarily in a representation that leaves the center intact. To be more precise for
\be
k ~~\mbox{mod}~~ N \neq 0,
\ee
a 't Hooft loop is a generalized Wilson loop in a representation that is sensitive to the center.
Therefore in such cases a 't Hooft loop can be detected by another 't Hooft loop.
In fact, a 't Hooft loop will pick up a non-trivial phase when going around another
't Hooft loop.
What this means is that $t(x)$ is an anyon. It is an easy task
to find its fractional statistics. By construction a 't Hooft loop creates a singular
gauge transformation with $U(2\pi)=e^{2\pi i/N}$. A generalized (or ordinary) Wilson loop,
 in the $\mbox{sym}k$ representation that winds once
around that loop will pick up a phase $(U(2\pi))^k=e^{2\pi ki/N}$ and hence the  fractional
statistics associated with  the 't Hooft operator is
\be\label{11q}
\delta ~~\mbox{mod}~~ \pi = \pi \frac{k}{N},
\ee
where the factor of $1/2$ is due to the fact that the statistics is determined to be
 one half the phase we get by winding once and we mod by $\pi$, since in determining
 the statistics that way there is a sign ambiguity (see footnote \ref{fo}).

In we take $g\rightarrow \infty$  the relevant piece of the action is  the CS term and the generalized Wilson
loop becomes just an ordinary Wilson loop. Thus eq.(\ref{11q}) should agree with Witten's results
on Wilson loop in pure CS theories \cite{Witten:88}. For example he found that the fractional
statistics of a Wilson loop in the fundamental representation is $(N^2-1)/2N(N+k)$. For $k=1$
this should agree with (\ref{11q}) since for $k=1$ the 't Hooft loop is a generalized
Wilson loop in the fundamental.
Indeed it does.
For $k=2$ the 't Hooft loop is a generalized Wilson loop in the symmetric piece of $N \otimes N$.
For that  representation Witten found $(N^2+N-2)/N(N+k)$ that for $k=2$
agrees with eq.(\ref{11q}).
To avoid  fractional statistics at all (including Fermions)
$k/2N$ must be an integer.  This agrees with
 \cite{ms} where is was argued, based on the relation
 between pure CS and WZW,  that for  $SO(3)=SU(2)/Z_2$ theory to
 avoid the analog of a fractional statistics in WZW theories $k/4$ must be an integer.

There is an important case for which the 't Hooft loop is not equivalent to a generalized
Wilson loop in some representation. When $k=0$, that is pure YM theory, eq.(\ref{uui}) makes no
sense and we have to take a few steps back to realize that formally the 't Hooft loop can
be written in that case in the following way
\be
T(C)=\int Dg(\tau) \exp\left(\pi \sqrt{\frac{2(N-1)}{N}} i\oint_C dx^{\mu}\mbox{Tr}
[H_{N-1}(g(\tau)^{-1}F_{\mu}~g(\tau)] \right).
\ee
The absence of a kinetic term for $g(\tau)$ makes it  hard to put this integral
in a useful form. Nevertheless in the next section we shall see that  much can be said about
the effective action of the 't Hooft operator in that case.

\section{Effective description}

One might hope that the fact that the 't Hooft loop is equivalent to a generalized Wilson loop
can be used to make some precise statements, that can be  supported by solid calculations,
 about the infra-red dynamics as a function of $N$ and $k$. In the lack of any concrete progress
in that direction we discuss in this section some possible effective actions for the 't Hooft loop
and speculate about the infra-red dynamics. First we review and elaborate a bit on 't Hooft construction
of the effective action
for the pure Yang-Mills theory \cite{thooft} and then we consider the $k\neq 0$ case.

\subsection{$k=0$}

In the  ultra-violet the disorder operators plays no significant role
 as the
theory is well described in terms of the classical
Yang-Mills action. In  the infra red it is
conceivable that the disorder operator takes over in the sense  that
the main features of the theory in the
deep infra red are described in terms of an effective field
theory of the disorder operator. This is the case  in some exactly solvable
two dimensional models.

In \cite{thooft} 't Hooft suggested a  simple effective action for the disorder parameter
that yields condensation of the disorder parameter and admits confinement.
Consider   Yang Mills theory with no Chern Simons term and
 assume that other matter fields in the theory, if present,  are in a representation
of the gauge group that does not transform  under the center, so that the 't Hooft operator
is a local operator. In the absence of a CS term the 't Hooft field, $t(x)$,
 is  a chargeless scalar (for $k=0$ there is no fractional statistics).
Nevertheless it is a complex field as
(for $N>2$) one can distinguish between a $Z_N$ instanton  and an anti-instanton.
For example a non-trivial operator in the theory is $ \langle t(x) t^{\dag}(y) \rangle $
that describes the instanton anti-instanton
two point function. Since $\pi_1 (SU(N)/Z_N) = Z_N$ if we have
$N$ 't Hooft lines  they can be annihilated and
hence the  instanton charge is conserved modulo $N$.
Therefore, in the effective action for $t$ there must be a $t^N$ vertex.
The simplest effective Lagrangian  is
\be\label{xc}
{\cal L}_{eff} = \partial_{\mu} t \partial^{\mu} t^{\dag} -m^2 t t^{\dag}
 - V(t t^{\dag}) -i \lambda (t^N - t^{\dag N}).
\ee
The action is invariant under the $Z_N$ symmetry $t\rightarrow e^{2\pi i/N} t$.

The sign of the mass term is a dynamical question whose answer depends on  $N$ and
 the matter fields. Let us assume that it is such that $t$ condenses which in turn means that
the $Z_N$ symmetry is broken. The condensation of the disorder parameter should lead to
 confinement in the order parameter via the dual Meissner effect. If so then where are the
closed strings associated with the electric flux tubes?
Since a discrete symmetry is broken  there are domain wall solutions
that interpolate between two vacua and since we are in three dimensions the domain wall is
a string. The  way to see that these strings are indeed the electric flux tube associated
with the insertion of a Wilson loop is the following.
In the Hamiltonian formalism  a Wilson loop in the fundamental
does not commute with $t(x)$ if the Wilson loop winds around $x$.
That is simply the Hamiltonian formalism analog of the
statement that when a Wilson loop winds around a 't Hooft loop it picks up a $Z_N$ phase.
That phase can be interpreted, in the Hamiltonian formalism, as shifting $t$ by a
$Z_N$ phase inside the Wilson loop relative to the outside of the loop
which is exactly what the domain wall solution does. The glue ball
spectrum can be calculated from fluctuations of the domain wall strings.

The way these domain wall
closed strings emerge from the gauge theory is a bit different than in the string
theory case. Indeed, the closed strings are dual to Wilson loops as was argued on general grounds
 in \cite{pol} and demonstrated in the context of the AdS/CFT duality  in \cite{rey,malda},
but here the strings are not
the fundamental degrees of freedom in the dual description. Here the dual description is in terms
of the 't Hooft operator and the strings emerge as domain wall solutions. In particular, they propagate in the
same number of dimensions as the gauge fields.

What happens when we introduce quarks? We expect to see now open strings with a quark at
one end and an anti quark at the other end. How to see this at the level of the operators was explained by
't Hooft. Here we suggest a way to think about this at the level of the effective action.

When we have fields in the fundamental (or any other representation that transforms non-trivially
under the center of the gauge group) the 't Hooft loop can be detected by the quarks which
 means that $t(x)$ is no longer a local operator but it depends on the path.
For example $ \langle t(x) t^{\dag}(y) \rangle $ does not seem to make sense any more without
 specifying
the path from $x$ to $y$. Does this mean
that in the presence of quarks there is no local effective description in terms of the 't Hooft
operator?
In other words, should we conclude that  the effective action ${\cal L}_{eff}$ cannot be modified
in a local fashion
that captures the effect
 of the quarks?

Not necessarily.
The fact that an operator like $ \langle t(x) t^{\dag}(y) \rangle $ depends on the path between $x$ and $y$
and not only on the end points is very familiar from gauge theories. There, a meaningful
electron positron operator will take the form $\phi(x)\exp(i\int_x^y dx^{\mu} A_{\mu})
\phi^{\dag}(y)$. Clearly, that operator depends on the path. What this is suggesting is that
from the point of view of the disorder operator the quarks are like gauge fields. That is,
the dual effect of adding quarks to Yang Mills
is to add a gauge symmetry to ${\cal L}_{eff}$. That gauge
symmetry should
not be confused with the original $SU(N)$ gauge symmetry. In fact, as we show below, a $U(1)$
gauge symmetry is sufficient. On general grounds we expect the coupling of $t(x)$ to the new gauge
field to
increase as we decrease the quark's mass. The reason being that if we have a very heavy quark
then the fact that it is  sensitive to the 't Hooft loop have a small effect in the path integral. Hence
$ \langle t(x) t^{\dag}(y) \rangle $ depends weakly on the path. In the gauge theory analogy this means
that the coupling constant is small and hence the effect of the Wilson line between the
electron and the positron is also small.

$t$ is a complex scalar, thus the simplest modification is
 to leave it as it is   and to couple it to a
$U(1)$ gauge field, $\tilde{A}_{\mu}$. We also want
 to add a Maxwell term for the gauge field, so that
the modified action takes the form
\be\label{aq}
{\cal L}_{eff}=-\frac14 F^2+ D_{\mu} t D^{\mu} t^{\dag} -m^2 t t^{\dag}
 - V(t t^{\dag}) + i \lambda (t^N - t^{\dag N}), ~~~D_{\mu}=\partial_{\mu} +i \tilde{g}
\tilde{A}_{\mu}.
\ee
The last term in the action is not invariant under the gauge transformation. Thus
 the gauge transformation is only an approximation that becomes better and better
as we flow to the infra-red since (for $N>2$) the last term is an irrelevant term.
As we now show
 the approximate gauge symmetry is sufficient to yield the correct infra red dynamics.

If we set $\lambda$ to  zero then there is a dual Nielsen Olesen solution where $t$ gets
 expectation value
that at infinity goes to
\be\label{ji}
t(\th) = t_0 e^{i \th}.
\ee
The reason why a soliton with such a boundary condition
costs only  a finite amount of energy is that this boundary condition  can be obtained as
a gauge transformation
from $t(x)=t_0$ and so there is no divergence coming from spatial infinity.
When $\lambda> 0$ a configuration with boundary condition (\ref{ji}) will cost an infinite
amount of energy
since the last term creates  a sin-Gordon like potential in the $\th$ direction.
To wind once at infinity with a finite energy
we have to consider a more general asymptotic behavior
\be\label{iu}
t(\th,r)= t_0 e^{i f(\th,r)},~~~\mbox{with}~~f(0,r)=0,~~f(2\pi,r)=2\pi.
\ee
To minimize the effect of
 the sine-Gordon like potential $f(\th)$ has to "jump" from one minimum of $t^N=t_0^N$ to the other.
So that all together there are $N$ "jumps". For any finite $r$ these jumps are continuous in $\th$.
In the limit $r\rightarrow \infty$ the jumps are discontinuous in $\th$.
Normally, that would cost a large amount of energy from the
kinetic term of
$t$ but since for any $f(\th)$ eq.(\ref{ji}) can be obtained as a gauge transformation of $t=t_0$
we need not worry about infinities  coming from the kinetic terms.
We should be a bit more careful here. The gauge symmetry is only
an approximated gauge symmetry. This means that acting with a gauge transformation on $t=t_0$
the contribution coming from all terms in (\ref{aq}) but the last one
is zero no matter what $f$ is. The contribution of the last term depends on $f$ and with
the $f$ described above it is  finite.
Simply put, one can  construct a finite energy configuration
that winds once at infinity by appropriate choice of $f(\th, r)$.
The actual solution, which is quite hard to find  as it depends both on $r$ and $\th$,
will have lower energy than this finite energy configuration and so the
soliton solution we are after must exist. Since that
solution makes $N$ "jumps"
it should be associated with a baryon. The baryon mass scales like $\sim 1/\tilde{g}^2$ and as
 anticipated it decreases as we increase the dual coupling constant.

One might attempt to describe a single quark by taking only $1/N$ the winding of eq. (\ref{iu}),
\be\label{iu2}
t(\th,r)= t_0 e^{i f(\th,r)},~~~\mbox{with}~~f(0,r)=0,~~f(2\pi,r)=2\pi/N.
\ee
That is, however, not a consistent boundary condition as $t$ does not go back to itself,
 $t(0)\neq t(2\pi)$.
 This is expected since in the phase where the 't Hooft operator condenses one should not find
 an isolated quark.
On the other hand if there is another solution with the exact same jump in $t$ then the two
can be glued together
to cancel the discontinuity. This is exactly what the domain-wall string solution does and
 hence we conclude that  the domain-wall string can end on a quark. A heuristic form of the solution that describes
an open string with a quark on one end and an anti-quark on the other end is

\vspace{-0.59in}

\[
\parbox{6.0in}{
\begin{picture}(150,120)(0,0)
\put(185,60){$t=t_0$}
\put(-13,40){$t=t_0 e^{if(\th_1,r) /N}$}
\put(352,49){$t=t_0 e^{i(2\pi-f(\th_2,r)) /N}$}
\put(175,20){$t=t_0 e^{2\pi i/N}$}
\vspace{5mm} \hspace{20mm} \mbox{\epsfxsize=96mm \epsfbox{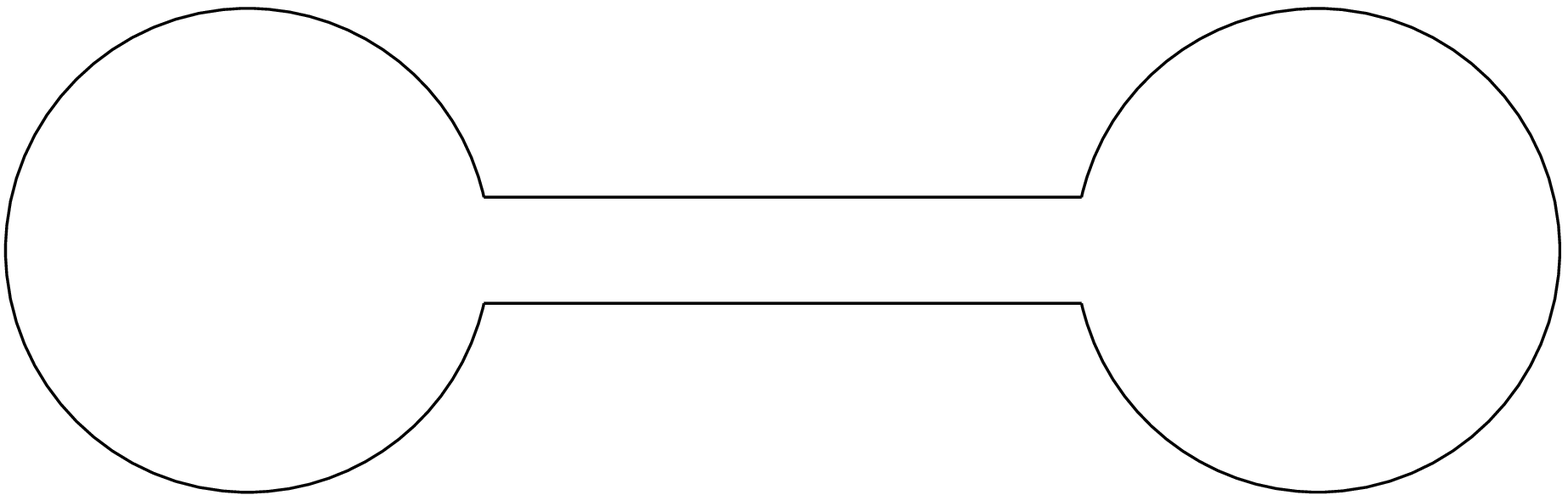}}
\end{picture}}
\]
Note that whether we have a quark or an anti-quark at the, say, left end is a matter of notation.
However, if we have a quark on one side we must have an anti-quark on the other side.
We see that also measons can be described in terms of soliton solutions. In particular, one should be able
to find a solution that corresponds to the Regge trajectory where the string tension is balanced by the centrifugal force.
Note that unlike in the Skyrme model \cite{sk,witten2}, where the baryons are solitons with respect to
 the meason fields, here the measons and baryons are on equal footing, both are described as soliton solutions.

If we add more than  one  quark the natural modification seems to involve several $U(1)$
gauge fields, one for each quark so that the gauge group is $U(1)^{N_f}$ and
 measons with different flavor can be constructed as solitons.
The idea that fields in the fundamental "transform" into gauge fields in the dual description is
reminiscent of Seiberg duality \cite{seiberg} where an $SU(N_c)$ gauge theory with $N_f$
quarks in the fundamental
is dual at the infra-red to a gauge theory with $SU(N_f - N_c)$ (with a small modification
this is correct also in three dimensions \cite{karch}) which implies that the dual gauge
group grows with $N_f$ much like in the case considered here.

\subsection{$k\neq 0$}

It is tempting to apply a similar approach to Chern-Simons-Yang-Mills theory.
However, as we will now see the effective description in terms of the disorder operator
in that case is much more complicated and it is less clear if it yields reasonable results.

For $k\neq 0$ the gauge fields acquire classically a mass $\sim g^2 |k |$ \cite{djt}.
Since the three dimensional analog of $\Lambda_{QCD}$ is $g^2 N$,
for $k\gg N$  the mass of the gauge fields is larger than
$\Lambda_{QCD}$ and the classical mass gap can be trusted to conclude that the
theory is in a Higgs like phase. The interesting question is what is happening
when $k$ is of the order or smaller than $N$.
In that case  $\Lambda_{QCD}$ is of the order or larger than the mass of the gauge fields.
This means that  the classical description breaks down and hence
the 't Hooft operator is expected to play an important role. The fact that
for $k\neq 0$  the 't Hooft loop is charged under the gauge group and it also has fractional statistics
suggests that,
at the level of effective description, things are much more complicated than in the pure
Yang-Mills case. In particular,  there is no hope  to find
a local effective action for $t(x)$ without including gauge fields as well.

For $k=1$ we saw in the previous section that $t$ is in the fundamental representation of the gauge group.
We also saw that it has fractional statistics $\pi /N$.
The simplest local effective action one can write that yield the correct fractional statistics is
\be\label{yc}
{\cal L}_{eff} =  CS+ D_{\mu} t D^{\mu} t^{\dag}  -V(t t^{\dag}),
\ee
where $ CS$ is the Chern-Simons term with $k=1$ that makes sure that $t$ has the correct fractional statistics.
Note that eventhough $t$ has fractional statistics it can condense as it is clear from (\ref{yc}).
 The reason being that
the fractional statistics comes from attaching a magnetic flux to charged particles. But there is no charge density
when $\langle t(x) \rangle = \mbox{const.}$ In other words, condensation of $t$ gives an anyonic plasma rather
than  anyonic gas.

This effective action is the simplest
also "conceptually". In the ultra-violet we describe the system via the classical Chern-Simons-Yang-Mills action.
Semi-classically the CS term is relevant compared to the YM term and so by {\em  assumption}
 it is kept  in ${\cal L}_{eff}$.
The YM term is replaced, much like in the pure Yang-Mills case, by the disorder terms.
Therefore, by  assuming
that form of the effective action in a sense we have already assumed that nothing dramatic is
going to happen in the infra-red. That is,  the theory is in the Higgs phase.
Indeed, since $t$ is charged  its
 condensation  implies screening.
To be more precise the smallest mass for the gauge field one finds when $t$ condences scales like $\sim |~t_0|^2$
\cite{pr},  and so we are in the Higgs phase.

For $k>1$ we found that $t$ is  in the $\mbox{sym}k$ representation of the gauge group so it has charge $k$
under the center of the gauge group. This suggests that condensation of $t$ leads to a phase in which all
Wilson loops in the $N^l$ representation for $l<k$  admit an area law while a Wilson loop
 in the $N^k$  representation is screened.
For example, if $k=N=2$ then
$t$ is in the adjoint and its condensation implies screening of adjoint charges, but
 much like in \cite{polyakov76},  instantonic effects induces an area law for a Wilson loop
 in the fundamental.

\section{Discussion}

In some theories in two dimensions  the order disorder map is a
powerful tool to study non-perturbative aspects.
In the 2D Ising model for example, the simplest description, in which it is most transparent
that the theory is dual to a free fermion, is in terms of a field that involves both
the order and the disorder parameter.
In that respect it is certainly encouraging  that in Chern-Simons-Yang-Mills theory
the disorder operator, the 't Hooft loop,  is equivalent
 to a generalized Wilson line that can be described in terms of the order parameter (the
connection) and its derivatives.
So far, however, we did not find a way to invert this observation into a precise
progress in the understanding of non-perturbative aspects of gauge theories in 3D.
For example, it would be very nice to actually show that the 't Hooft operator condenses (or not).
It might be that the findings in this paper should be combined
with some other technique to yield concrete progress.
One such possibility is to take $N$ to be large. The motivation here
 is  not so much  the connection with string
theory, as much as to  tune the fractional statistics of the
't Hooft operator to be  almost fermionic
or almost bosonic by an appropriate choice of $k$.
In the case of anyonic gas that was proven to be useful
to improve the validity of the mean field theory approximation \cite{ww}.
 It should be interesting to see
 whether it is helpful in the present case as well.

A different approach that is worthwhile to explore is to add supersymmetry.
The disorder operator discussed here
 is most relevant for mirror symmetry in three dimensions \cite{Intriligator}.
Its crucial role in mirror symmetry was emphasized  in \cite{Aharony}
for  $U(1)$ and $SU(N)$ broken down to $U(1)^{N-1}$. In fact, in the $U(1)$ case it can
be used to make some
 non-trivial relations between the mirror theories away from the deep infra-red \cite{ks}
 (for recent papers see \cite{b,bb}).
We hope that the results of the present paper will shead  some light on
 mirror symmetry when the gauge group is not broken.
In particular, with enough supersymmetry there are scalars
 in the adjoint. In that case, one can consider a more general
connection that contains covariant derivatives of the scalars ontop of the dual field strength
in such a way that some supersymmetry is preserved. It seems reasonable  that such an
operator will play a role in  the non-Abelian mirror symmetry transformation.

\newpage


\nl {\bf Acknowledgments}

\nl I thank N. Drukker, N. Seiberg, H. Verlinde and E. Witten for discussions at various stages of this work.
This material is based upon work supported by the National Science
Foundation under Grant No. PHY 9802484.
Any opinions, findings, and conclusions or recommendations expressed in
this material are those of the author and do not necessarily reflect
the views of the National Science Foundation.

\end{document}